\documentclass[10pt,conference]{IEEEtran}
\IEEEoverridecommandlockouts
\usepackage{cite}
\usepackage{amsmath,amssymb,amsthm,amsfonts,bm,graphicx,hyperref,mathtools,physics,qcircuit,tikz,upgreek}
\usepackage[linesnumbered,ruled,vlined]{algorithm2e}
\usepackage{graphicx}
\usepackage{textcomp}
\usepackage{xcolor}
\usepackage{caption}
\usepackage{subcaption}

\def\BibTeX{{\rm B\kern-.05em{\sc i\kern-.025em b}\kern-.08em
    T\kern-.1667em\lower.7ex\hbox{E}\kern-.125emX}}

\newtheorem{lemma}{Lemma}

\makeatletter
\newcommand{\linebreakand}{
  \end{@IEEEauthorhalign}
  \hfill\mbox{}\par
  \mbox{}\hfill\begin{@IEEEauthorhalign}
}
\makeatother

\begin{document}

\title{$k$-Contextuality as a Heuristic for Memory Separations in Learning\\\thanks{* These authors contributed equally}
\thanks{$^\dagger$ Corresponding Authors}}
\author{
    \IEEEauthorblockN{Mariesa H. Teo*$^\dagger$}
    \IEEEauthorblockA{\textit{Pritzker School of Molecular Engineering}\\
    \textit{University of Chicago}\\
    Chicago, IL \\
    mhteo@uchicago.edu}
\and
    \IEEEauthorblockN{Willers Yang*$^\dagger$}
    \IEEEauthorblockA{\textit{Dept. of Computer Science} \\
    \textit{University of Chicago}\\
    Chicago, IL \\
    willers@uchicago.edu}
\and
    \IEEEauthorblockN{James Sud}
    \IEEEauthorblockA{\textit{Dept. of Computer Science} \\
    \textit{University of Chicago}\\
    Chicago, IL \\
    jsud@uchicago.edu}
\linebreakand
    \IEEEauthorblockN{Teague Tomesh}
    \IEEEauthorblockA{\textit{Infleqtion} \\
    Chicago, IL\\
    teague.tomesh@infleqtion.com}
\and
    \IEEEauthorblockN{Frederic T. Chong}
    \IEEEauthorblockA{\textit{Dept. of Computer Science} \\
    \textit{University of Chicago}\\
    Chicago, IL\\chong@cs.uchicago.edu}
\and
    \IEEEauthorblockN{Eric R.\ Anschuetz}
    \IEEEauthorblockA{
    \textit{IQIM \& Burke Institute} \\
    \textit{Caltech}\\
    Pasadena, CA\\
    eans@caltech.edu}
}
\maketitle

\begin{abstract} 
Classical machine learning models struggle with learning and prediction tasks on data sets exhibiting long-range correlations. Previously, the existence of a long-range correlational structure known as contextuality was shown to inhibit efficient classical machine learning representations of certain quantum-inspired sequential distributions. Here, we define a new quantifier of contextuality we call strong $k$-contextuality, and prove that \emph{any} translation task exhibiting strong $k$-contextuality is unable to be represented to finite relative entropy by a classical streaming model with fewer than $k$ latent states. Importantly, this correlation measure does not induce a similar resource lower bound for quantum generative models. Using this theory as motivation, we develop efficient algorithms which estimate our new measure of contextuality in sequential data, and empirically show that this estimate is a good predictor for the difference in performance of resource-constrained classical and quantum Bayesian networks in modeling the data. Strong $k$-contextuality thus emerges as a measure to help identify problems that are difficult for classical computers, but may not be for quantum computers.%In this paper, we develop algorithms to estimate how strongly $k$-contextual a data set is, and study empirically how this estimate correlates with the performance of quantum and classical hidden Markov models on synthetic and practically-relevant data sets exhibiting strong $k$-contextuality.
\end{abstract}
\begin{IEEEkeywords}
contextuality,
quantum machine learning,
quantum advantage
\end{IEEEkeywords}

\section{Introduction}
To justify the use of quantum computers over classical computers in any given problem domain, one must show the existence of useful problems that are relatively easy for a quantum computer to solve while still being relatively hard for a classical computer to solve. We investigate the latter requirement in the setting of machine learning. In particular, we study the limitations of classical generative models through the lens of \emph{contextuality}, a correlational structure introduced in quantum foundations theory, and investigate how the presence of contextuality is related to the resource requirements and the performance of classical hidden Markov models relative to their quantum counterparts.

\begin{figure}
    \centering
    \includegraphics[width=0.9\linewidth]{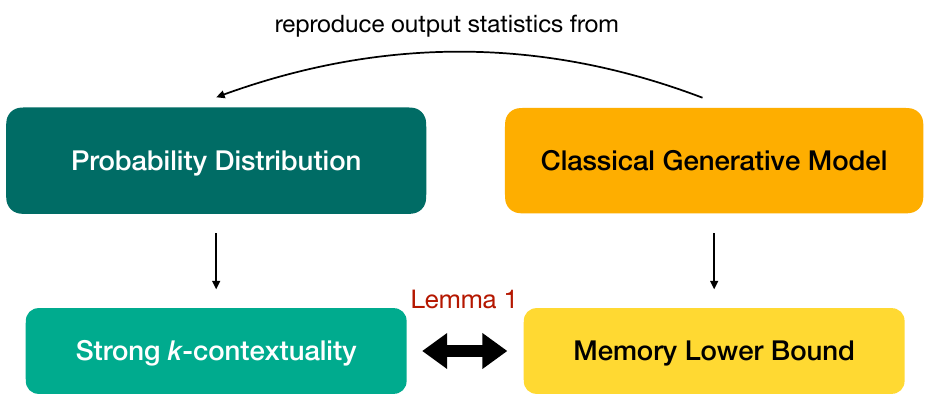}
    \caption{Strong $k$-contextuality implies a memory lower bound on a classical generative model representing some family of probability distributions to finite relative entropy.}
    \label{fig:bigpicture}
\end{figure}

The intuition to study the implications of contextuality on machine learning (ML) stems from the fact that classical ML models can have difficulty capturing the correlations present in certain complex probability distributions---such as those associated with the measurement statistics of quantum systems\cite{arute2019quantum, ZHU2022240, doi:10.1126/science.abe8770}---and measures of contextuality capture these complex correlations. Indeed, several studies demonstrating separations between quantum-enhanced ML and classical ML attribute the separation to contextuality \cite{gao2021enhancing, anschuetzgao2022, anschuetz_arbitrary_2024}. 

We extend the sheaf framework for contextuality \cite{Abramsky_2011} by introducing a quantity we call \emph{strong $k$-contextuality}. 
%of the minimum memory resources required for a classical machine learning model to represent a given probability distribution
We prove that modeling a distribution exhibiting strong $k$-contextuality requires a minimum amount of memory for a classical computer. When $k$ is large, representing such distributions becomes infeasible for classical generative models. Importantly, this lower bound does not hold for ML models with a quantum memory. We thus propose strong $k$-contextuality as a measure to identify sequence learning problems whose classical memory requirements scale intractably, narrowing down the search space for learning problems that have the potential to demonstrate quantum advantage. 
\begin{figure}
    
    \includegraphics[width=0.85\linewidth]{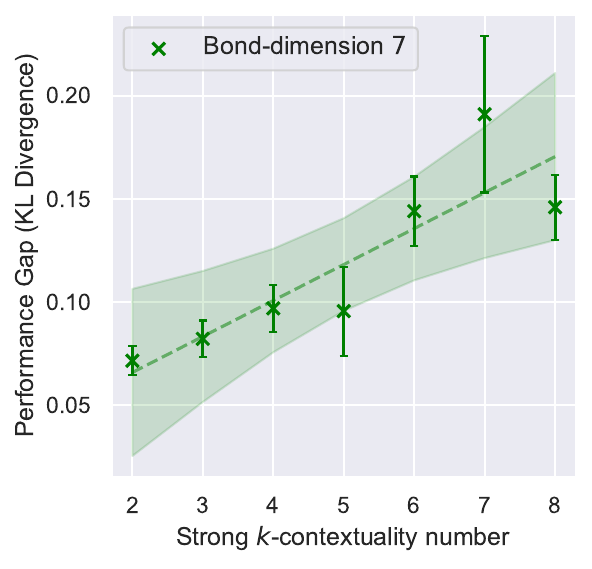}
    \caption{Presence of strong $k$-contextuality in empirical data sets can indicate separations between classical and quantum learning models. Here we show an example of the performance gap between classical and quantum HMMs trained on synthetic random distributions, where the gap increases with the larger strong $k$-contextuality number. We return to this result in Sec. \ref{sec:example}.}
    \label{fig:strongkcont-hero}
\end{figure}

This paper is structured as follows: We begin in Sec.~\ref{sec:background} by reviewing concepts central to understanding this paper -- we introduce hidden Markov models, which form the basis of many of our theoretical and empirical studies, in  Sec.~\ref{sec:hmm_review}, and briefly overview the sheaf-theoretic notion of contextuality in Sec.~\ref{sec:contextuality}. In Sec. \ref{sec:strong_k_contextuality}, we define strong $k$-contextuality and use it to prove a memory lower bound for classical hidden Markov models. We develop methods for finding how strongly $k$-contextual a data set is in Sec. \ref{sec:algorithms}, and benchmark them in Sec. \ref{sec:benchmarking}. Finally, in Sec. \ref{sec:example}, we apply our heuristic algorithms to estimate the strong $k$-contextuality of several sequential learning tasks, demonstrating a correlation between the estimated strong $k$-contextuality and a separation in the resource requirements of classical and quantum generative models in representing the distribution to some fixed target error.

\section{Background}\label{sec:background}

\subsection{Hidden Markov models}\label{sec:hmm_review}

% Generalizability of Bayesian Network
% basis enhancement as the minimal quantum extension

We first briefly introduce generative models, and then delve more deeply into hidden Markov models. 

Broadly, generative learning seeks to approximate a probability distribution $P$ learned via a set of training data $\mathcal{D} = \{x_1,...,x_n\}$ drawn from $P$. While for fully general $P$ learning the distribution efficiently is intractable, in practice $P$ can often be assumed to be described by few underlying parameters. A \emph{generative model} attempts to approximate $P$ by fitting a parameterized family of distributions $\{p_\theta\}_{\theta\in\Theta}$ to $\mathcal{D}$; that is, one attempts to find a choice of $\hat{\Theta}$ such that $p_{\hat{\Theta}}$ is close in statistical distance to $P$. Here we use the relative entropy (sometimes referred to as KL divergence) as a measure of statistical distance:
\begin{equation}
    D(P||Q) := \sum_{x}P(x)\log \frac{P(x)}{Q(x)}.
\end{equation}
By convention, we define $p\log q = 0$ when $p = 0$, and $p\log q = \infty$ when $p>q= 0$. 

A hidden Markov model (HMM)~\cite{pearl1985bayesian,10.1214/aoms/1177699147} is a canonical example of a generative model that attempts to represent $P$ as emissions from an underlying Markov chain on a cardinality-$m$ state space $\varLambda$. HMMs also include more standard machine learning models as special cases, including recurrent neural networks (RNNs)~\cite{Hopfield2554}, long short-term memory (LSTM) networks~\cite{10.1162/neco.1997.9.8.1735}, transformer decoders~\cite{10.5555/3295222.3295349}, and any other autoregressive models implemented at any finite precision~\cite{9635256}.

More explicitly, an HMM begins in a state $\lambda_1\in\varLambda$, and receives tokens $x_1,\ldots,x_\ell$ from an input space $X$ sequentially. Upon receipt of the token $x_i$, the HMM samples from a distribution $q\left(o_i\mid\lambda_i,x_i\right)$ over output tokens $o_i\in O$, where $\lambda_i\in\varLambda$, and updates its latent representation via sampling from $t\left(\lambda_{i+1}\mid\lambda_i,x_i,o_i\right)$. The final distribution is:
\begin{equation}\label{eq:hmm_prob}
    \begin{aligned}
        p\left(\bm{o}\mid\lambda_1,\bm{x}\right)=&q\left(o_\ell\mid\lambda_\ell,x_\ell\right)\\
        &\times\prod_{i=1}^{\ell-1} t\left(\lambda_{i+1}\mid\lambda_i,x_i,o_i\right)q\left(o_i\mid\lambda_i,x_i\right).
    \end{aligned}
\end{equation}
It is easy to see that $m=\left\lvert\varLambda\right\rvert$ controls the expressivity of a given HMM, as trivially the class of HMMs with $\left\lvert\varLambda\right\rvert=m-1$ is contained in those with $\left\lvert\varLambda\right\rvert=m$. Given an HMM, we also define the marginal distribution over latent variables after an input sequence $\bm{x}$:
\begin{equation}\label{eq:hmm_single_output}
    t_\leq\left(\lambda\mid\lambda_1,\bm{x}\right):=\sum_{\substack{o_1,\ldots,o_{i-1}\\\lambda_2,\ldots,\lambda_{i-1}}}\prod_{j=1}^{i-1} t\left(\lambda_{j+1}\mid\lambda_j,x_j,o_j\right)q\left(o_j\mid\lambda_j,x_j\right),
\end{equation}
which we will refer to later.

\subsection{The contextuality framework} \label{sec:contextuality}
We now briefly review the concept of contextuality. In the context of quantum mechanics, contextuality refers to the property that measurement statistics cannot be predicted by any hidden variable theory---in other words, the outcomes of measurements do not correspond to merely revealing preexisting assignments of values to those observables.

Though the study of contextuality was initiated by the quantum foundations community, the concept has since been generalized to settings unrelated to quantum mechanics~\cite{Abramsky_2011,acin2015combinatorial,PhysRevLett.115.150401}. We begin by reviewing this generalized definition of contextuality, basing our presentation on the sheaf-theoretic construction due to Ref.~\cite{Abramsky_2011}. We will later define a new measure of contextuality we call \emph{strong $k$-contextuality} in this sheaf-theoretic setting, and show that it is related to memory lower-bounds for classical simulation methods.

The sheaf-theoretic framework for contextuality describes a particular correlation structure in conditional probability distributions~\cite{Abramsky_2011}. In particular, one considers a set $\left\{e_C\right\}_C$ of conditional distributions:
\begin{equation}
    e_C\left(\bm{o}\mid\bm{x}\right),
\end{equation}
where the $x_i\in\bm{x}$ are assumed to belong to some fixed set $X$, and the $o_i\in\bm{o}$ to some fixed set $O$. Here, $C$ labels a \emph{context}, a subset of $X$ such that all $x_i\in C\subseteq X$. We use $\mathcal{M}\subseteq 2^X$ to denote the set of all contexts and assume it is a cover of $X$ (i.e., $\bigcup_{C\in\mathcal{M}}C=X$). The $e_C$ are also assumed to be \emph{consistent} in that their marginals agree wherever their associated contexts intersect. In particular, for all $C,C'\in\mathcal{M}$, $\bm{x}\subseteq C\cap C'$, and $\bm{o}\in O^{\left\lvert\bm{x}\right\rvert}$, we have:
\begin{equation}\label{eq:consist_rel}
    e_C\left(\bm{o}\mid\bm{x}\right)=e_{C'}\left(\bm{o}\mid\bm{x}\right).
\end{equation}
We call this collection $\left(X,O,\mathcal{M},\left\{e_C\right\}\right)$ an \emph{empirical model}. We say an empirical model is \emph{contextual} if there exists no distribution $p\left(\bm{o}\mid\bm{x}\right)$ independent of $C$ satisfying:
\begin{equation}
    p\left(\bm{o}\mid\bm{x}\right)=e_C\left(\bm{o}\mid\bm{x}\right)
\end{equation}
for all $C\in\mathcal{M}$, $\bm{x}\subseteq C$, and $\bm{o}\in O^{\left\lvert\bm{x}\right\rvert}$. In other words, contextual empirical models \emph{require} multiple distributions to fully characterize them. We say an empirical model without this property is \emph{noncontextual}.

% In the sheaf-framework for contextuality \cite{Abramsky_2011}, the scenario studied is as follows: Some number of agents, each of whom can choose one measurement from a collection of possible measurements, conduct their chosen measurements simultaneously. Each measurement has its own corresponding outcomes, and some joint outcome of all the simultaneous measurements is observed. This event can be repeated, and the relative frequencies of the possible joint outcomes can be described by some probability distribution, given the subset of chosen measurements. For each possible subset of chosen measurements, there is some probability distribution of outcomes, and this family of probability distributions is called an empirical model. If this empirical model cannot be reproduced by a hidden variable model, it is \textit{contextual}.

While this definition seems ad hoc, it allows one to capture settings typically encountered in generative modeling. For instance, consider a setting where one is interested in translating a long stream of text from English to Spanish. One might consider a data set composed of two streams:
\begin{enumerate}
    \item The zoo got a new bat. That bat is black.
    \item He bought a new baseball bat. That bat is black.
\end{enumerate}
While the second sentence in both streams is identical in the source language, in the target language they differ:
\begin{enumerate}
    \item El zoológico tiene un murciélago nuevo. Ese murciélago es negro.
    \item Compró un bate de béisbol nuevo. Ese bate es negro.
\end{enumerate}
We claim we can write this translation task as learning an empirical model. Here, $X$ is the set of English vocabulary, and $O$ the set of Spanish vocabulary. Given this data set, $\mathcal{M}$ is a set of two contexts:
\begin{equation}
    \begin{aligned}
        \mathcal{M}=&\left\{C_1,C_2\right\}\\
        :=&\left\{\left\{\text{The},\text{zoo},\text{got},\text{a},\text{new},\text{bat},\text{that},\text{is},\text{black}\right\},\right.\\
        &\left.\left\{\text{He},\text{bought},\text{a},\text{new},\text{baseball},\text{bat},\text{that},\text{is},\text{black}\right\}\right\}.
    \end{aligned}
\end{equation}
Finally, $\left\{e_{C_1},e_{C_2}\right\}$ are distributions which marginalize to distributions with nonintersecting supports:
\begin{equation}\label{eq:cont_example}
    \begin{aligned}
        \operatorname{supp}&\left(e_{C_1}\left(\text{That},\text{bat},\text{is},\text{black}\right)\right)\\
        &\cap\operatorname{supp}\left(e_{C_2}\left(\text{That},\text{bat},\text{is},\text{black}\right)\right)\\
        =&\left\{\left(\text{Ese},\text{murciélago},\text{es},\text{negro}\right)\right\}\\
        &\cap\left\{\left(\text{Ese},\text{bate},\text{es},\text{negro}\right)\right\}\\
        =&\varnothing.
    \end{aligned}
\end{equation}
Here, we use the notation $\operatorname{supp}\left(e_C\left(\bm{x}\right)\right)$ to denote the support of the conditional distribution $e_C$ when conditioned on $\bm{x}$. By construction, this empirical model is contextual.

Indeed, this example of contextuality is also an example of \emph{strong contextuality}~\cite{Abramsky_2011}. Strong contextuality is a statement that not only does there not exist a distribution $p\left(\bm{o}\mid\bm{x}\right)$ independent of $C$ which correctly marginalizes to the $\left\{e_C\right\}_C$, but also there exists no $p\left(\bm{o}\mid\bm{x}\right)$ with even the correct \emph{support}. More formally, defining the set:
\begin{equation}
    \begin{aligned}\label{eq:s_e_m_def}
        S_e^\mathcal{M}=&\left\{p:\forall C\in\mathcal{M}\text{ and }\forall \bm{x}\subseteq C,\right.\\
        &\left.\operatorname{supp}\left(p\left(\bm{x}\right)\right)\subseteq\operatorname{supp}\left(e_C\left(\bm{x}\right)\right)\right\},
    \end{aligned}
\end{equation}
we say an empirical model is strongly contextual if:
\begin{equation}
    S_e^\mathcal{M}=\varnothing.
\end{equation}
Finally, we say that two contexts $C_i,C_j\in\mathcal{M}$ are \emph{compatible} if:
\begin{equation}
    S_e^{\left\{C_i\right\}}\cap S_e^{\left\{C_j\right\}}\neq\varnothing,
\end{equation}
or equivalently if:
\begin{equation}
    \forall \bm{x}\subseteq C_i\cap C_j,\operatorname{supp}\left(e_{C_i}\left(\bm{x}\right)\right)\cap\operatorname{supp}\left(e_{C_j}\left(\bm{x}\right)\right)\neq\varnothing.
\end{equation}

% A statement of contextuality relates to whether or not we are able to represent this empirical model using a \textit{global section}, a section over all of $X$. Specifically, we say an empirical model is \textit{strongly contextual} if the set

% \begin{equation} \label{eq:strongcontextuality}
%     S_e^\mathcal{M} := \{s \in \mathcal{E}(X):\forall C \in \mathcal{M}, s_C \in \text{supp}(e_{C})\}
% \end{equation}

% is empty. In words, an empirical model is strongly contextual if there is no global value assignment that is consistent with the true measurement distribution when restricting to contexts. 

\section{Strong \texorpdfstring{$k$}{k}-contextuality} \label{sec:strong_k_contextuality}
\subsection{Definition}
\begin{figure}
    \centering
    \includegraphics[width=\linewidth]{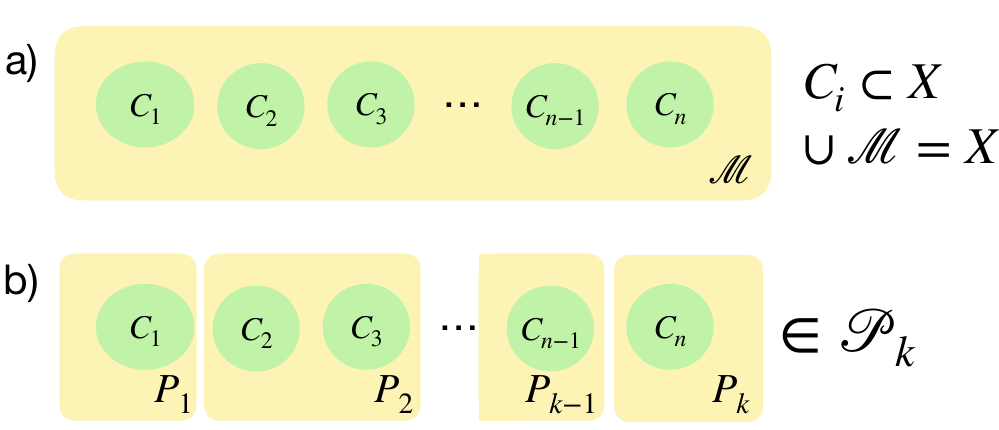}
    \caption{Overview of the contextuality framework. a) A measurement cover $\mathcal{M}$ is a collection of contexts $\{C_i\}$, each of which is a subset of commuting measurements in $X$. The union over all the contexts covers all the measurements in $X$, ie. $\bigcup_{C \in \mathcal{M}} C = X$. For each context $C_i\in\mathcal{M}$, an empirical model assigns some probability distribution $e_{C_i}$ over the joint outcomes of the elements of $C_i$. b) When considering strong $k$-contextuality, we consider every way to partition $\mathcal{M}$ into $\leq k$ subsets $\mathcal{M}_i$. The sets of all $k$-partitions is denoted $\mathcal{P}_k$.}
    \label{fig:strongcontextualitydef}
\end{figure}

We now extend the definition of strong contextuality to \textit{strong $k$-contextuality}. We let $\mathcal{P}_k\left(\mathcal{M}\right)$ denote the sets of $k$-partitions of $\mathcal{M}$ (see Fig. \ref{fig:strongcontextualitydef}), i.e.,
\begin{equation}
    \left\{P_1,\ldots,P_k\right\}=P\in\mathcal{P}_k\left(\mathcal{M}\right)\iff\bigsqcup_{i=1}^k P_i=\mathcal{M},
\end{equation}
and define the number:
\begin{equation}
    N_e^k:=\sum_{P\in\mathcal{P}_k\left(\mathcal{M}\right)}\prod_{i=1}^k\left\lvert S_e^{P_i}\right\rvert,
\end{equation}
where $S_e^{P_i}$ is as in Eq.~\eqref{eq:s_e_m_def}. We say that an empirical model is strongly $k$-contextual if:
\begin{equation}
    N_e^k=0.
\end{equation}
Informally, an empirical model is strongly $k$-contextual if not only does there not exist a single distribution $p$ correctly marginalizing to the empirical model over all contexts, but also this property is robust under partitioning the set of contexts into any $k$ subsets. By construction, the $k=1$ case is equivalent to the definition of strong contextuality. Furthermore, $N_e^k\leq N_e^{k+1}$ as:
\begin{equation}
    \begin{aligned}
        N_e^{k+1}=&\sum_{P\in\mathcal{P}_{k+1}\left(\mathcal{M}\right)}\prod_{i=1}^{k+1}\left\lvert S_e^{P_i}\right\rvert\\
        =&\sum_{P\in\mathcal{P}_{k+1}\left(\mathcal{M}\right):P_{k+1}=\varnothing}\prod_{i=1}^k\left\lvert S_e^{P_i}\right\rvert\\
        +&\sum_{P\in\mathcal{P}_{k+1}\left(\mathcal{M}\right):P_{k+1}\neq\varnothing}\prod_{i=1}^{k+1}\left\lvert S_e^{P_i}\right\rvert\\
        =&N_e^k+\sum_{P\in\mathcal{P}_{k+1}\left(\mathcal{M}\right):P_{k+1}\neq\varnothing}\prod_{i=1}^{k+1}\left\lvert S_e^{P_i}\right\rvert\\
        \geq&N_e^k,
    \end{aligned}
\end{equation}
so empirical models that are strongly $k+1$-contextual are also strongly $k$-contextual. Finally, note that no model is strongly $\left\lvert\mathcal{M}\right\rvert$-contextual, as trivially the distribution $e_C$ satisfies Eq.~\eqref{eq:s_e_m_def} when $\left\lvert P_i\right\rvert=1$. Thus, we will say that an empirical model has \emph{contextuality number $k$} when $k+1$ is the smallest integer such that the model is not strongly $\left(k+1\right)$-contextual.

\subsection{Proof of classical memory lower bound} \label{sec:memorylb}
We now discuss the importance of strong $k$-contextuality. We claim that if an empirical model $\left(X,O,\mathcal{M},\left\{e_C\right\}\right)$ is strongly $k$-contextual, there exists a memory lower bound for any HMM simulation of it that achieves a finite relative entropy with the true distribution.

Specifically, we assume a setting where one is given tokens $x_1,\ldots,x_\ell\in C$ sequentially. The task is to sample from a distribution $p_C\left(\bm{o}\mid\bm{x}\right)$ such that the relative entropy of $p_C$ from $e_C$ is finite:
\begin{equation}\label{eq:rel_ent}
    D\left(p_C||e_C\right)=\sum_{\bm{x},\bm{o}}p_C\left(\bm{o}\mid\bm{x}\right)\ln\left(\frac{p_C\left(\bm{o}\mid\bm{x}\right)}{e_C\left(\bm{o}\mid\bm{x}\right)}\right)<\infty.
\end{equation}
This problem is well-defined as, if $x_1,\ldots,x_\ell$ belong to multiple contexts $\left\{C_i\right\}$, the consistency relation of Eq.~\eqref{eq:consist_rel} ensures the associated $e_{C_i}\left(\bm{o}\mid\bm{x}\right)$ agree.

Our main result in this section lower bounds the cardinality of $\varLambda$ for any HMM achieving a finite relative entropy in simulating $\left(X,O,\mathcal{M},\left\{e_C\right\}\right)$ in terms of the strong $k$-contextuality of the empirical model.
\begin{lemma}[Strong $k$-contextuality yields lower bounds on classical simulation]
    Let $\left(X,O,\mathcal{M},\left\{e_C\right\}\right)$ be an empirical model that is strongly $k-1$-contextual. Any HMM simulating $\left(X,O,\mathcal{M},\left\{e_C\right\}\right)$ to any finite relative entropy must have at least $k$ hidden states.
\end{lemma}

Intuitively, this lower bound stems from the fact that an HMM cannot represent two contexts using the same latent state if the empirical model over these contexts cannot be consistently represented by a fixed distribution. The minimal number of distributions needed to cover all the contexts then translates to the number of latent states needed for the model to represent the empirical model. A more rigorous proof is as follows.

\begin{proof}
    Consider an HMM $p\left(\bm{o}\mid\lambda_1,\bm{x}\right)$ achieving a finite relative entropy with the empirical model. Recall from Eq.~\eqref{eq:hmm_single_output} that $t_\leq\left(\lambda\mid\lambda_1,C\right)$ is the distribution of the latent state of the HMM after receiving entries from $C$ in sequence when beginning in some arbitrary $\lambda_1\in\varLambda$; we fix $\lambda_1$ and leave its dependence implicit. Consider an arbitrary ordering $\left(\lambda_1,\ldots,\lambda_m\right)$ of the elements of $\varLambda$, where recall $m=\left\lvert\varLambda\right\rvert$. We iteratively define for each latent state $\lambda_i$ the set of contexts (that have not yet been covered by $P_{\lambda_1},..., P_{\lambda_{i-1}}$) that $\lambda_i$ supports:
    \begin{equation}
        P_{\lambda_i}:=\left\{C\not\in\bigsqcup_{j=1}^{i-1}P_{\lambda_j}:\lambda_i\in\operatorname{supp}\left(t_\leq\left(C\right)\right)\right\}
    \end{equation}
    such that:
    \begin{equation}
        \bigsqcup_{i=1}^m P_{\lambda_i}=\mathcal{M}.
    \end{equation}
    Now, define:
    \begin{equation}
        p_{\lambda_i}\left(\bm{o}\mid\bm{x}\right):=p\left(\bm{o}\mid\lambda_i,\bm{x}\right),
    \end{equation}
    where $p\left(\bm{o}\mid\lambda_i,\bm{x}\right)$ is the HMM distribution as defined in Eq.~\eqref{eq:hmm_prob}. We claim that, for all $C\in P_{\lambda_i}$ and $\bm{x}\subseteq C$,
    \begin{equation}\label{eq:p_e_c_rel}
        \operatorname{supp}\left(p_{\lambda_i}\left(\bm{x}\right)\right)\subseteq\operatorname{supp}\left(e_C\left(\bm{x}\right)\right).
    \end{equation}
    To see this, we first define:
    \begin{equation}
        p_C\left(\bm{o}\mid\bm{x}\right):=\sum_{j=1}^m p\left(\bm{o}\mid\lambda_j,\bm{x}\right)t_\leq\left(\lambda_j\mid C\right)
    \end{equation}
    for conciseness. Note that:
    \begin{equation}\label{eq:p_c_supp_first_rel}
        \operatorname{supp}\left(p_{\lambda_i}\left(\bm{x}\right)\right)\subseteq\operatorname{supp}\left(p_C\left(\bm{x}\right)\right),
    \end{equation}
    which follows due to $C\in P_{\lambda_i}$ implying that $\lambda_i$ is in the support of $t_\leq\left(C\right)$. As the HMM is assumed to achieve a finite relative entropy, it must also be the case that:
    \begin{equation}
        -\sum_{\bm{o}}p_C\left(\bm{o}\mid\bm{x}\right)\ln\left(e_C\left(\bm{o}\mid\bm{x}\right)\right)<\infty,
    \end{equation}
    which in particular means:
    \begin{equation}
        \operatorname{supp}\left(p_C\left(\bm{x}\right)\right)\subseteq\operatorname{supp}\left(e_C\left(\bm{x}\right)\right).
    \end{equation}
    This together with Eq.~\eqref{eq:p_c_supp_first_rel} implies Eq.~\eqref{eq:p_e_c_rel}. In the language of Eq.~\eqref{eq:s_e_m_def},
    \begin{equation}
        p_{\lambda_i}\in S_e^{P_{\lambda_i}}.
    \end{equation}
    In particular,
    \begin{equation}
        N_e^m \geq\prod_{i=1}^m\left\lvert S_e^{P_{\lambda_i}}\right\rvert\geq 1,
    \end{equation}
    and the empirical model is not strongly $m$-contextual. As the empirical model is strongly $k-1$-contextual by assumption, this implies that $\left\lvert\varLambda\right\rvert=m>k-1$, yielding the desired result.
\end{proof}

Hence, given the contextuality number of an empirical model, we can lower bound the resources required for a classical machine learning model to learn about it. Intriguingly, strong $k$-contextuality provides no equivalent memory lower bounds for quantum models. A similar observation led to a provable quantum-classical memory separation in simulating Clifford circuits in \cite{gao2021enhancing}.

\section{Methods for finding strong \texorpdfstring{$k$}{k}-contextuality}\label{sec:algorithms}
Perhaps unsurprisingly, determining an empirical model's contextuality number is itself a computationally intensive problem because this measure quantifies the resources required for a classical ML model to solve the corresponding learning problem.
This is evident from the definition of strong $k$-contextuality: an empirical model is strongly $k$-contextual if, for \textit{every} partition that splits the contexts into $k$ subsets, there is no valid $k$-sized collection of value assignments, each corresponding to one subset. 
To ensure that there is no such valid set of value assignments, one must check every possible $k$-partitioning of contexts, and for each choice, check for a consistent assignment within each of the $k$ subsets. 

We devise a brute force search algorithm to find the exact solution, as specified in Alg. \ref{alg:exact}.%\era{is it right to call this greedy? I would call it a brute force search}

\begin{algorithm}[h!]
  \caption{Brute force search algorithm for finding contextuality number exactly}
  \label{alg:exact}
  \DontPrintSemicolon
  
  \SetKwProg{Fn}{def}{\string:}{}
  \SetKwInOut{Input}{Input}
  \SetKwInOut{Output}{Output}
  \SetKwFor{For}{for}{:}{}
  \SetKwComment{Comment}{\# }{}
  \SetKwIF{If}{ElseIf}{Else}{if}{:}{elif}{else:}{}
  \SetKw{In}{in}
  \SetKw{Break}{break}
  \SetKw{Or}{or}
  \BlankLine
  
  \Input{empirical model $e_C$ with measurement cover $\mathcal{M}= \{C_i\}$}
  \Output{contextuality number $k$}
  \BlankLine
    $\mathcal{P} \gets \text{all possible permutations of } \mathcal{M}$\;
    $N \gets \{\}$\;
    \For{permutation $p$ \In $\mathcal{P}$}{
        $\mathcal{S} \gets \{\}$ \;
        \For{$C_i$ \In $p$}{
            \For{$S$ \In $\mathcal{S}$}{
                \If{$C_i$ is compatible with the other contexts in $S$}{
                    add $C_i$ to $S$\;
                    \Break\;
                }
            }
            \If{$C_i$ is not compatible with any of the existing subsets \Or $\mathcal{S} = \emptyset$}{
                add $\{C_i\}$ to $\mathcal{S}$ \#{ create a new subset}\;
            }
        }
        add $k=|\mathcal{S}|$ to $N$ \#{ number of subsets needed to find a valid set of assignments for $p$}\;
    }
    \Return{min($N$)$-1$}

\end{algorithm}
\begin{algorithm} [t]
  \caption{Hypergraph coloring algorithm for estimating contextuality number}
  \label{alg:hypergraph}
  \DontPrintSemicolon
  
  \SetKwProg{Fn}{def}{\string:}{}
  \SetKwInOut{Input}{Input}
  \SetKwInOut{Output}{Output}
  \SetKwFor{For}{for}{:}{}
  \SetKwComment{Comment}{\# }{}
  \SetKwIF{If}{ElseIf}{Else}{if}{:}{elif}{else:}{}
  \SetKw{In}{in}
  \SetKw{Break}{break}
  \SetKw{Or}{or}
  \BlankLine
  
  \Input{empirical model $e_C$ with measurement cover $\mathcal{M}= \{C_i\}$, sparsity $d$}
  \Output{contextuality number $k$}
  
  \#{ Step 1: Create incompatibility hypergraph}\;
  $\mathcal{G} \gets \text{Hypergraph with }|e|\text{ nodes}$\;
  \For{i = 1 to $d$}{
    \For{each subset of contexts of size $i$}{
        \If{contexts in the subset are incompatible}{
            Add hyperedge of size $i$\;
        }
    }
  }

  \#{ Step 2: Coloring approximation algorithm \cite{plociennik2008approximation}}\;
  $\mathcal{S} \gets \{\}$ \#{Each group in $\mathcal{S}$ represents one color}\;
  \For{node $v$ \In $\mathcal{G}$}{
    \For{color group $S$ \In $\mathcal{S}$}{
        \If{$S \cup \{v\}$ contains no hyperedges of $\mathcal{G}$}{
            $S \gets S \cup \{v\}$ \#{Add $v$ to the first valid color group}\;
        }
        }
    \If{$v$ was not added to any color group \Or $\mathcal{S} = \emptyset$}{
        add $S = \{v\}$ to $\mathcal{S}$ \#{ Create a new color group}\;
    }
  }
  \Return{$|\mathcal{S}|-1$}

\end{algorithm}
However, at large system sizes, finding the exact contextuality number becomes computationally intractable. For an empirical model with $n$ contexts, the algorithm must search through all of the $n!$ possible orderings of the $n$ contexts; for each of these orderings, it must check compatibility between each of the $n$ contexts with up to $(n-1)$ subsets, and within each subset, check compatibility with up to $(n-1)$ contexts. The time complexity of the greedy method thus scales in the worst case as $\mathcal{O}(n! \times n^3)$.

We thus develop two methods to approximate the contextuality number, and discuss their expected time complexitites. The first, a greedy heuristic that follows naturally from the above brute force search algorithm, can be used for any empirical model. The second, an algorithm for approximating the chromatic number of a hypergraph, applies only to cases where the empirical model is sparse.

\subsubsection{Greedy heuristic for general empirical models}

To avoid having to iterate through every possible ordering of the contexts in the empirical model, the greedy heuristic randomly selects a fixed number of order permutations. Hence, in line 1 of Alg. \ref{alg:exact}, instead of $\mathcal{P}$ being the set of all possible permutations of the empirical model $e$, a set number of these permutations are sampled. The rest of the algorithm follows Alg. \ref{alg:exact}. The time complexity for this approximation method thus circumvents the factorial, scaling as $\mathcal{O}(n^3)$.

\subsubsection{Contextuality as a hypergraph for sparse models} 
In the case where an empirical model is $s$-sparse, meaning there are $\leq s$ possible outcomes for each context, our key insight is showing that the problem of finding a valid global value assignment reduces to a hypergraph coloring problem (see Appendix \ref{apx:coloring}). This is a problem which has already been studied in previous literature, and for which approximation algorithms have been developed~\cite{plociennik2008approximation}.

By constructing a rank-$\left(d+1\right)$ %\era{do you mean the cardinality of each hyperedge? I think that's called the ``rank'' of the hypergraph, i.e. I'd say ``by constructing a rank-$\left(d+1\right)$ hypergraph} 
hypergraph $G = (V,E)$ where each node $v\in V$ corresponds to a context, and each hyperedge $e\in E$ corresponds to a mutually compatible set of contexts, we can then use existing approximations for the hypergraph coloring problem \cite{plociennik2008approximation} to estimate the contextuality number. The full hypergraph coloring algorithm for estimating contextuality number can be found in Alg. \ref{alg:hypergraph}, where the runtime of the algorithm is upperbounded by \begin{align}
    T 
    &= n_{colors} \times n_{edges}\times n_{nodes/edge} \\
    &\leq n\times \sum_{i = 2}^s {n\choose i}\times s\label{eq:runtime_color}\\
    &\leq \mathcal{O}(n^{s+2}).
\end{align}

Note that the runtime upper bound is loose when $s>\frac n2$.

%\era{you talk about mutual compatibility in a bunch of places in this section, and I noticed there wasn't a definition. I added a definition of what I think you mean to the end of Sec. 2, double check it's actually what you mean}

We compare the worst-case time complexities for the two different approximation methods to the exact algorithm in Table \ref{tab:complexity}. %Since the greedy heuristic randomly samples ordering permutations of contexts, there is no guarantee on how closely it approximates the true contextuality number. The approximation ratio for the coloring algorithm is determined in Ref. \cite{plociennik2008approximation}.For the coloring algorithm, since the cardinality of the hyperedges depends on sparsity $s$, the time complexity scales with sparsity: $\mathcal{O}(n^{s+2})$ \cite{plociennik2008approximation}. 

\begin{table} [h!]
    \centering
    \begin{tabular}{c|cc}
         &  Time Complexity \\ \hline
         Brute force search (exact) & $\mathcal{O}(n! \times n^3)$ \\
         Greedy (approximation) & $\mathcal{O}$($n^3$) \\
         Coloring (approximation) & $\mathcal{O}$($n^{s+2}$) \\
    \end{tabular}
    \caption{Time complexities of exact and approximation methods for finding the contextuality number of an empirical model with $n$ contexts and sparsity $s$. %Note the runtime upper bound is loose when $s>\frac n2$ for the coloring algorithm. %era{I'm assuming the approximation guarantee is additive here? if it's multiplicative then it's larger than the total number of contexts asymptotically. so assuming it's additive, this definition differs from the ``overparameterization ratio'' we use in the main text. for this reason I would say something like ``the coloring algorithm is guaranteed to return a context number $k_{\text{est}}\leq k^\ast+\epsilon$, where $k^\ast$ is the true contextuality number and $\epsilon\leq\mathcal{O}$($n \log\log(n)^2 / \log(n)^2$''}.
    }
    \label{tab:complexity}
\end{table}

Note that these approximations always overestimate contextuality number. By definition, it is impossible to partition a strongly $k$-contextual empirical model into $\leq k$ subsets of mutually compatible contexts. These methods thus always find a partitioning with $>k$ subsets, efficiently providing an upper bound on the contextuality number of a given data set.%\era{I would also mention that these are upper bounds for the contextuality number, which is pretty neat! (though doesn't directly say anything about memory lower bounds since it's in the wrong direction)}

\section{Benchmarking} \label{sec:benchmarking}

To benchmark the performance of our approximation methods, we tested them on two types of empirical models: (1) Random models, whose contextuality number could be verified numerically, and (2) GHZ models, the empirical models from measuring an $n$-particle GHZ state in different bases, whose contextuality number can be verified analytically. Note that we have overloaded the variable $n$ to parameterize the size of the empirical model for each case.

\subsection{Random models}\label{ssec:bm-random}
We tested the performance of our two approximation methods on random models, where we controlled both the size and the sparsity of the empirical models, setting the number of contexts, the number of measurements in a context, and the number of possible outcomes to the same value $n$. For verification, we ran the exact brute force search algorithm specified in Alg. \ref{alg:exact}, which limited the system size we were able to scale to. For each $n$ between $3$ and $10$, and $s$ from $1$ to $n$, we generate $500$ random empirical models with number of contexts $n$ and sparsity $s$, comparing the estimates of contextuality number from one iteration of each approximation method. 

\begin{figure} [h]
  \centering
  \includegraphics[width=0.84\linewidth]{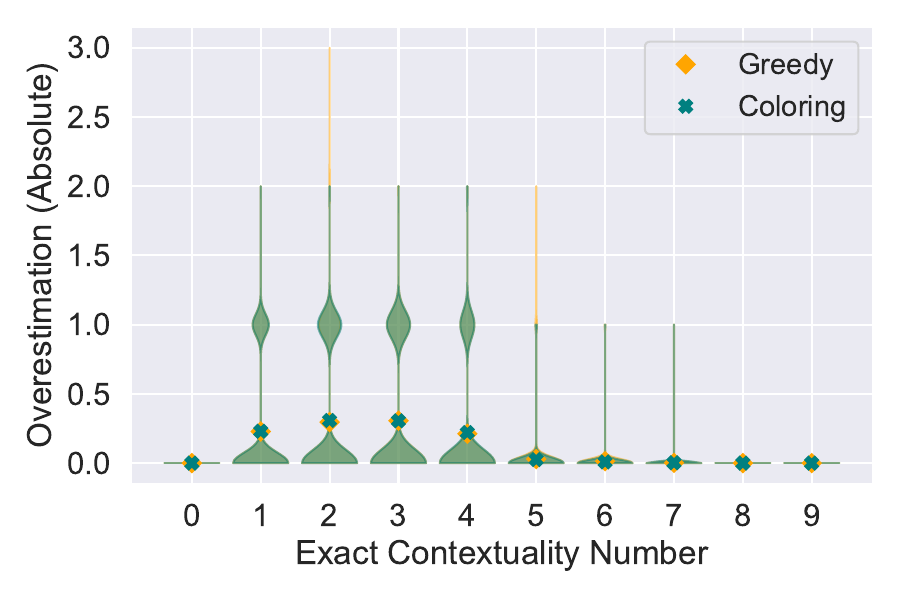}
  \caption{Performance of approximation methods for randomly generated empirical models, where shaded area represents the density of the distribution. As shown by the overlap in the shaded regions, the approximation methods often output the same contextuality number estimate. In most cases, both methods either find the correct value or overestimate the contextuality number by 1. %(a) We group the data by number of contexts $n$, and see that the overestimation ratio for both methods increases (gets worse) as the number of contexts increases. (b) We group the data by sparsity $s$, and see that both approximation methods struggle more for random models with middling sparsity values. The shaded regions here indicate one standard deviation away from the mean.%\era{I think the title for fig. 3(b) is incorrect? currently it's the same as for fig. 3(a). also, I would explain what the shaded region represents (I assume the standard deviation over some number of trials? is this at a fixed random instance though and randomizing the algorithms, or are you randomizing over both and looking at the standard deviation that way?)}
  }
  \label{fig:approxratio}
\end{figure}

\begin{figure} [h]
    \centering
    \includegraphics[width=0.84\linewidth]{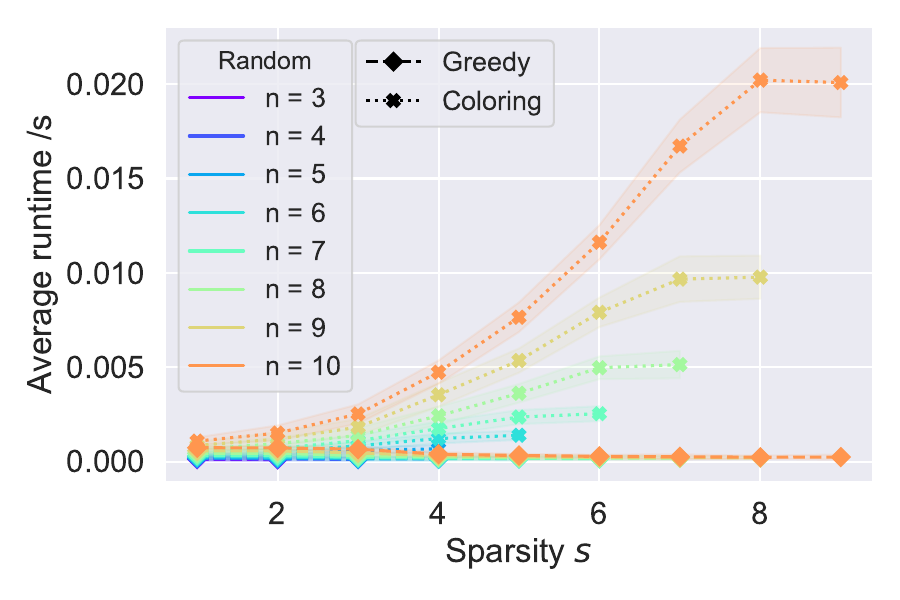}
    \caption{Runtimes of approximation methods for random models with $n$ contexts and $\leq s$ possible outcomes for each context. Time complexity of the coloring algorithm follows Equation~\ref{eq:runtime_color}.}
    \label{fig:algtimes}
\end{figure}

\begin{figure} [h!]
    \centering
    \includegraphics[width=0.84\linewidth]{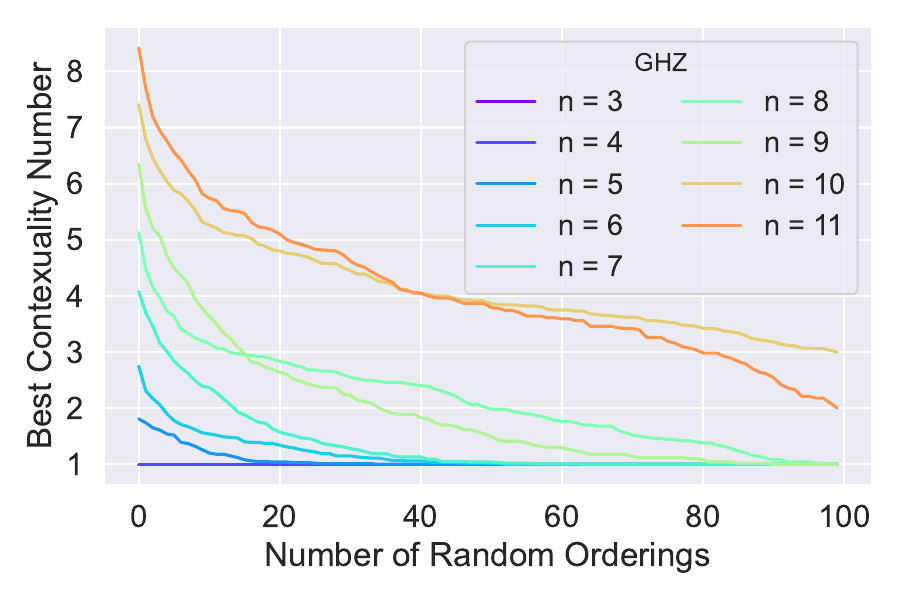}
    \caption{The greedy heuristic converges to the correct answer ($k=1$) within 100 random ordering permutations for GHZ models with up to $n=9$ particles ($2^9 > 500$ contexts). Larger system sizes required a larger number of random orderings.}
    \label{fig:ghz}
\end{figure}

\begin{figure*}
    \centering
    \includegraphics[height=0.25\textheight]{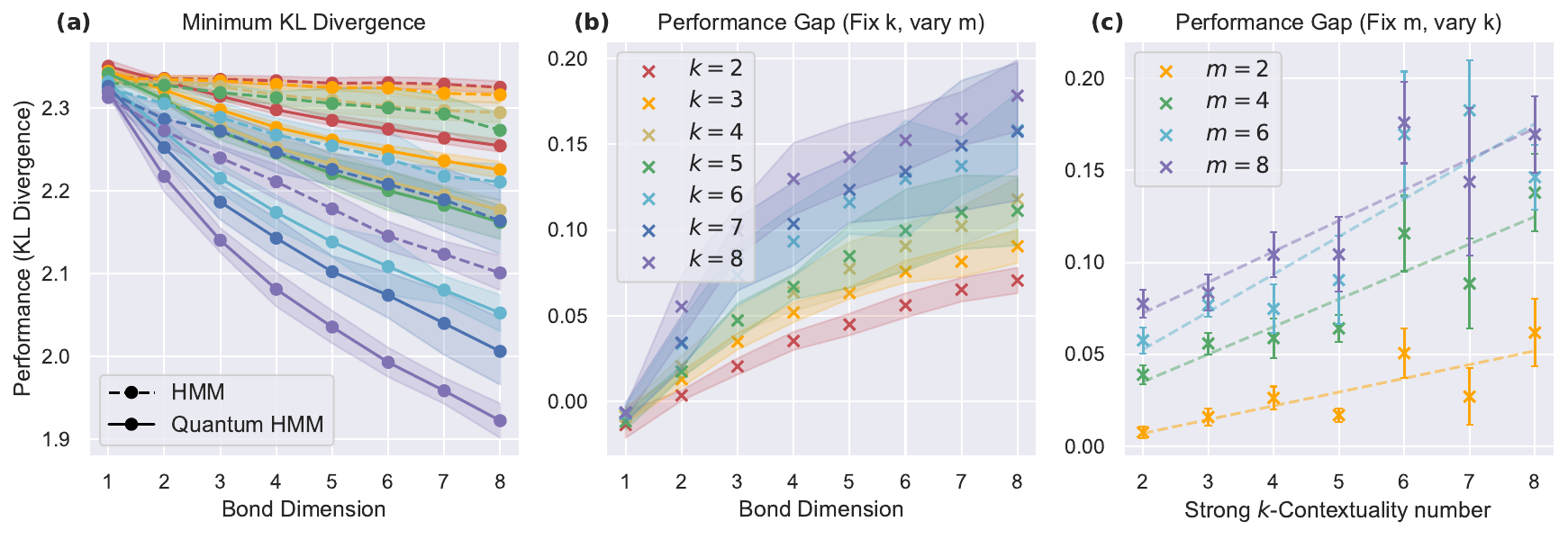}
    \caption{Performance of classical and quantum hidden Markov models on contextual random empirical models. (a) The KL divergence achieved by classical and quantum models (lower is better) for empirical models with contextuality number $k$ from 2 to 8 as bond dimension $m$ is increased from 2 to 8. The gap between the two models increases with both $m$ and $k$. This widening gap is plotted as a function of bond dimension in (b), and as a function of contextuality number in (c).}
    \label{fig:qhmm-random}
\end{figure*}

Our results, summarized in Fig. \ref{fig:approxratio}, show that both the coloring and greedy approximation methods perform similarly in terms of their accuracy, with both methods often producing the same estimate for the contextuality number. In most cases, the approximation methods either find the correct contextuality number, or overestimate by 1. Though the system sizes we are able to verify using the exact algorithm is limited, these results show promising performance for small random models with contextuality numbers between 0 and 9.   %Here, we have quantified performance in terms of what we call the overestimation ratio---the output of the approximation method divided by the exact solution. Note that we always expect the approximation ratio to be greater than or equal to $1$, since the approximation heuristics give an upper bound on the contextuality number.  

Though the approximation methods perform similarly for small models in terms of accuracy, the sparsity constraints of the coloring algorithm begin to show through in time complexity. As shown in Fig. \ref{fig:algtimes}, the runtime of the coloring method for an $s$-sparse empirical model increases with $s$ as expected from Equation~\ref{eq:runtime_color}, with the marginal increase scaling according to ${n\choose s}\times ns$. In further benchmarking, this scaling proved to be intractable at larger system sizes and sparsities.

%This time complexity follows from our understanding of the coloring algorithm: When constructing the hypergraph, the degree of the hyperedges up to which we need to check incompatibility for scales with sparsity. 

%Since the greedy algorithm is faster with similar performance to coloring, one obvious question is in which situations the coloring algorithm would be useful. Our motivation for developing the coloring algorithm was its guaranteed approximation ratio. As the greedy algorithm has no such performance guarantee, it is possible that for a model with a large number of contexts and sufficiently low sparsity, the coloring algorithm might outperform the greedy algorithm. However, as the size of the test models here is small, any benefit from the performance guarantee of the coloring algorithm is not obvious at this scale.

\subsection{GHZ models}\label{ssec:bm-ghz}
The second type of empirical model we used to benchmark our algorithms were GHZ models, which describe the measurement statistics of measuring each particle in an $n$-particle GHZ state $\frac{1}{\sqrt{2}}(|00...0\rangle + |11...1\rangle)$ in the Pauli $x$ or $y$ basis. This model has contextuality number $1$ for all $n$ \cite{Abramsky_2011}.

Note that the GHZ empirical model is not sparse---indeed, during our testing, this model was computationally intractable for the hypergraph coloring algorithm even for small $n$. We thus present results here just for the approximate greedy heuristic, showcasing the convergence of the algorithm with increasing system size. 

When we ran the greedy approximation for random models, we allowed just one iteration---i.e., we ran lines 2-13 of Alg. \ref{alg:exact} with just one ordering permutation of the empirical model. With the GHZ model, we wanted to see how the number of permutations we allowed affected the convergence of the algorithm to the correct answer. Fig. \ref{fig:ghz} shows the algorithm's best estimate of the contextuality number with an increasing number of permutations, for GHZ models of up to 11 particles ($2^{11} = 2048$ contexts). 

For GHZ models with up to $9$ particles, the greedy algorithm converged to the correct answer within the $100$ different permutations. For larger system sizes, the overestimation ratio would benefit from a greater number of permutations. Nevertheless, these results show that the greedy algorithm is able to reach the correct answer for a non-sparse empirical model of up to just over $500$ contexts.

\section{Strong k-contextuality in learning problems}\label{sec:example}

In this section, we will study the connection between strong $k$-contextuality and the performance gap between a model of classical and quantum machine learning. Focusing on both synthetic and practically relevant data sets, we apply the approximation methods developed in Sec. \ref{sec:algorithms} to empirically estimate their strong $k$-contextuality, train classical hidden Markov models and (simulations of) quantum models on them, and compare the performance of the models. We do this for a wide range of model sizes. Our numerical experiments reveal a widening performance gap between classical and quantum models as we increase the contextuality of the input distribution, while we also observe in two cases that the performance gap remains insensitive to the dimension of the inputs.

The quantum model we choose to compare the classical model against is known as a \emph{basis-enhanced HMM}~\cite{gao2021enhancing}, which we here refer to as a quantum HMM (QHMM). QHMMs generalize HMMs by---given a coherent implementation of an HMM~\cite{PhysRevA.89.062315}---lifting the measurement basis to any choice of local basis. These quantum models can be equivalently expressed as a tensor network, where the bond dimension $m$ of the network equals the cardinality of the latent space of the underlying HMM; that is, QHMMs at bond dimension $m$ directly generalize classical HMMs with $m$ hidden states. Due to the connection between HMMS, QHMMs, and tensor network simulation, we use the terms ``latent dimension'', ``number of hidden nodes/states'', and ``bond dimension'' interchangeably in this paper.

\begin{figure}
    \centering
    \includegraphics[width=0.5\textwidth]{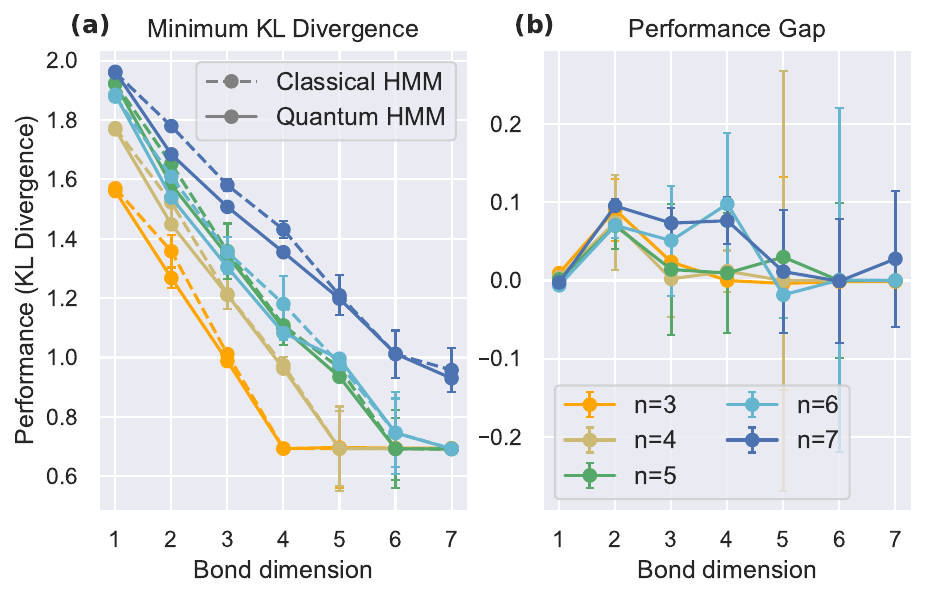}
    \caption{Performance of classical and quantum hidden Markov models on measurement outcomes of the $n$-particle GHZ state, which is always strongly 1-contextual, ie. the minimum number of hidden nodes needed to represent this empirical model is 2. (a) The KL divergence achieved by the two models for $n=3$ to $7$ with increasing bond dimension. (b) As expected, the performance gap between the two models does not persist beyond small bond dimension.}
    \label{fig:qhmm-ghz}
\end{figure}

In what follows, we train the classical HMMs using the Baum--Welch algorithm~\cite{baum1970maximization} and the quantum models using Riemannian gradient descent on a tensor network simulation implemented in the ITensor software library~\cite{itensor}. We refer the reader to \cite{gao2021enhancing} for further details of the QHMM simulations. The results are reported in Fig.s~\ref{fig:qhmm-random}-~\ref{fig:liklihood}. 

% Describe model -> data process for each and show the performance gaps
% point out correlation

\subsection{Random models} 
% can vary contextuality
% small so can exactly find contextuality
% widening gap that correlates with contextuality
% explanation?
We begin by studying the performance of classical and quantum HMMs on training data sampled from random empirical models with varying contextuality. Although contrived, the contextuality numbers of small random empirical models are flexible and easy to determine, which makes them an ideal candidate for studying the performance trends in relation to contextuality. 

The random empirical model is described in Sec.~\ref{ssec:bm-random}. We sample random empirical models of size $n=8$ and find their contextuality number exactly.  Each model encodes a conditional outcome distribution of measuring $n$ observables depending on the context; i.e., $\mathbb P [X_i = O_j|C_k]$ is assigned at random for each observable $X_i$, outcome $O_j$, and context $C_k$. We classify the random models based on their strong contextuality number $k$, and obtain $100$ random empirical models with $k=1,...,8$.

Then, for each model, we generate training data from the empirical models by randomly sampling the outcomes of each observable according to the conditional distribution. We train the HMM and QHMM on the resulting dataset and quantify the performance by the lowest KL divergence achieved. Then, we report the mean of the minimum KL divergences achieved by classical and quantum HMMs for context number $k=1,...,8$ in Fig.~\ref{fig:qhmm-random}, and show that there is a clear trend that the gap between the performance of classical and quantum HMMs widens as we increase contextuality and model size. We see that although our resource lower bounds using strong k-contextuality do not predict a widening performance gap in this parameter regime, we still observe a strong signal based on the contextuality of the probability distribution, which we cannot explain using existing techniques.

\subsection{GHZ models} 
% larger number of contexts
% can find the contextuality number analytically
% model performance is highest at d=2 -> contextual number = 1? 
Fig.~\ref{fig:qhmm-ghz} shows the classical and quantum HMMs trained on measurement outcomes of an $n$-particle GHZ state described in Sec.~\ref{ssec:bm-ghz}. While the size of the empirical model, which is $2^n$, grows exponentially with particle number, the contextuality number is always 1. In principle, an HMM may require as few as $2$ hidden nodes to represent the distribution within finite entropy. As expected, we observe a performance gap around a small bond dimension that does not scale with model size. 
\subsection{Promoter gene models}
% practical relevance
% show similar scaling to Random Models

% control--getting rid of dependence on token length
% control--getting rid of dependence on location

% behavior of classical models in higher bond dimensions
% explanation?

\begin{figure} 
    \centering
    \includegraphics[width=0.9\linewidth]{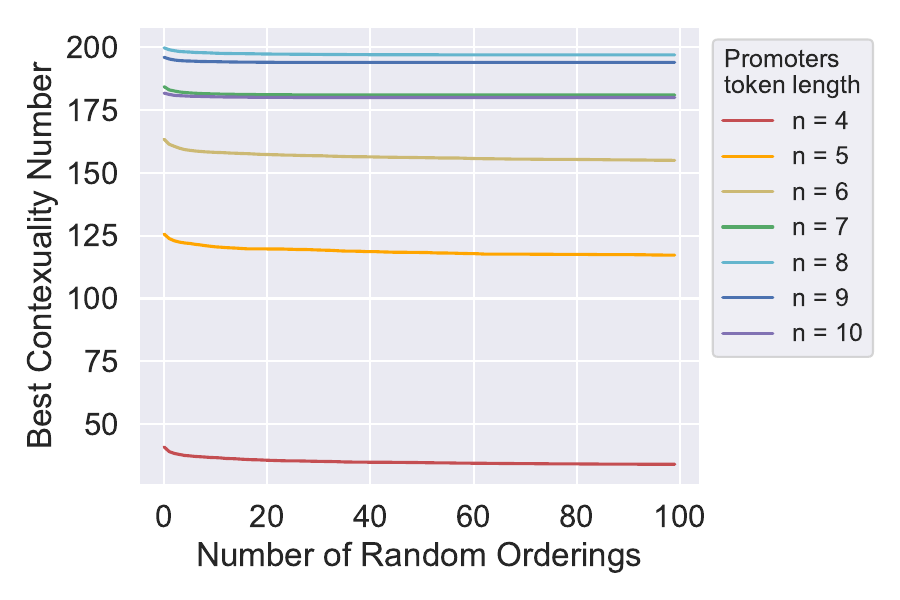}
    \caption{The convergence of the greedy heuristic to a contextuality number estimate for the promoter gene models within 100 random ordering permutations. The estimate of contextuality number increases with token length until $n=8$, and then starts to decrease for $n=9$ and $10$.}
    \label{fig:pmt_contextuality_estimates}
\end{figure}
\begin{figure*}
    \centering
    \includegraphics[height=0.37\textheight]{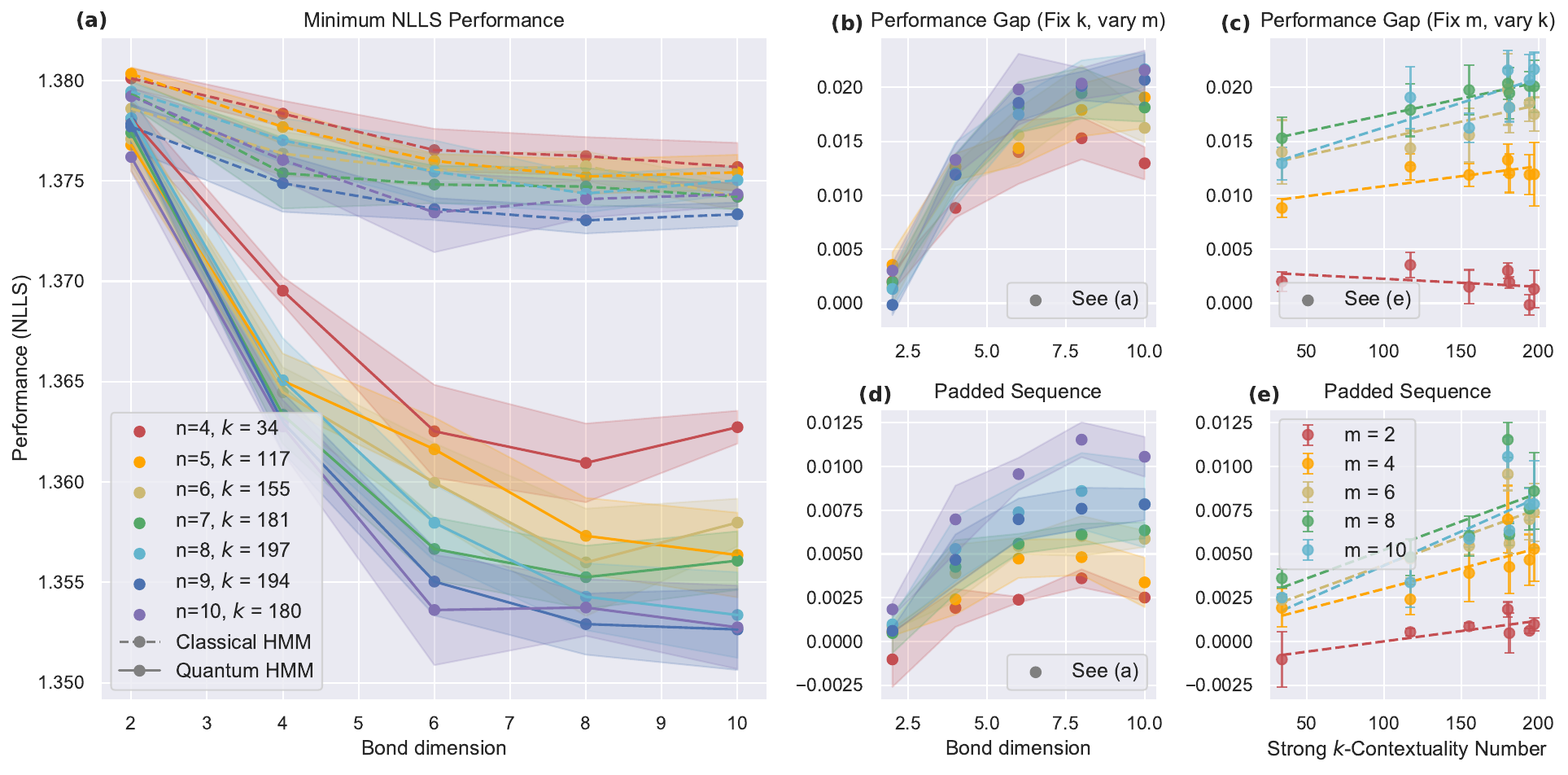}
    \caption{Performance of classical and quantum hidden Markov models on predicting promoter gene sequences of token length $n$. (a) The KL-divergence achieved by the two models for $n = 4$ to $10$, where the estimated contextuality number $k$ is computed using the greedy heuristic from Sec. \ref{sec:algorithms} as shown in Fig. \ref{fig:pmt_contextuality_estimates}. (b-e) The performance gap increases as we increase bond dimensions $m$ for all token lengths and contextuality numbers (b,d), and as contextuality number increases for bond dimensions $m>2$ (c,e). To control for token length, sequence examples in (d, e) are padded with randomly sampled tokens, so each data point has the same token length $n=15$. We observe clearer increasing trends with the padded sequences.}
    \label{fig:pmt-qhmm}
\end{figure*}
\begin{figure}
    \centering
    \includegraphics[width=\linewidth]{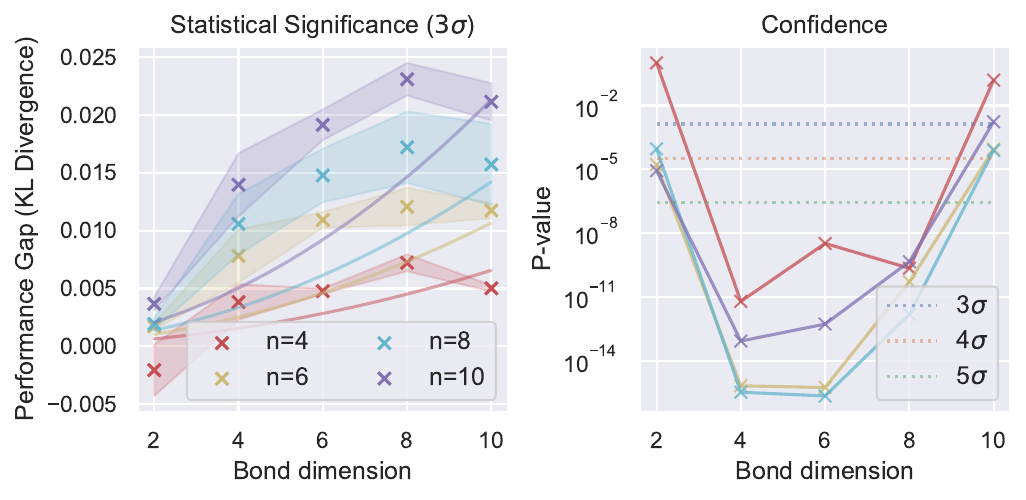}
    \caption{The likelihood-ratio test for the performance gap found in Fig.~\ref{fig:pmt-qhmm}. (a) The solid lines show the $\chi^2$ statistics (scaled by token lengths and data size) needed to reject the null hypothesis in a likelihood-ratio test with $3\sigma$ confidence. (b) The minimum significance level s.t. the null hypothesis is rejected. Rejecting the null with a lower $p$-value corresponds to a higher degree of confidence.}
    \label{fig:liklihood}
\end{figure}

Finally, we turn our attention to a practically relevant data set consisting of DNA sequences from promoter regions \cite{harley1990uci}.
Promoter regions are non-coding DNA segments that regulate gene expression by serving as binding sites for RNA polymerase and transcription factors, making their identification and characterization essential for understanding genetic disorders, drug development, and synthetic biology applications. From a machine learning perspective, promoter sequences represent an intriguing computational challenge due to the non-local correlations present within the human genome \cite{sanyal2012long, le2019classifying}. These characteristics make promoter gene sequences an attractive real-world data set to evaluate our $k$-contextuality framework.

We formulate the learning problem as token prediction, where given an $n$-length sub-segment of a promoter gene sequence, the goal is to predict the next $n$-length segment. As an empirical model, the first $n$ nucleotides are viewed as observables, and the following $n$ nucleotides are viewed as the measurement outcomes; hence, the joint distribution over all $2n$-length segments of a promoter gene is encoded in the empirical modal as a conditional distribution over $n$-length segments dependent on a distribution over the preceding $n$-length segments. We refer to $n$ as the size of the empirical model.

As seen in Fig.~\ref{fig:pmt_contextuality_estimates}, the estimated contextuality number of a size-$n$ empirical model increases with $n$ up to $n=8$, after which it decreases. We train the classical and quantum HMM on the data sampled from models of sizes $4,...10$ and show the performances in Fig.~\ref{fig:pmt-qhmm}. Similar to random models, we observe an increase in the performance gap as the contextuality number of the empirical model increases. 

While we observe a correlation between the performance gap and the contextuality number, the signal may be confounded with the increasing sequence lengths. To address this issue, we also study a transformation of the sequence samples where shorter sequences are artificially padded with randomly sampled nucleotides so all sequences have length $n = 15$. This transformation does not change the strong $k$-contextuality number of the underlying distribution, and the trend with performance gap persists, with small fluctuations that can be attributed to noise.

Finally, we perform a likelihood-ratio test to demonstrate the statistical significance of the performance gap in Fig.~\ref{fig:liklihood}. With $3\sigma$ confidence, we can \emph{reject} the null hypothesis that the empirical model can be better explained by the classical HMM for most parameters $n,k,m$. The smallest significance level at which the null hypothesis would be rejected also decreases as the sequence length and contextuality number increase, which means that the performance gap becomes more statistically significant as the contextuality number increases.

\section{Discussion}
%First, we define strong $k$-contextuality, a quantity which we prove provides a lower bound on the classical memory resources required for a broad class of generative models to represent a given probability distribution.  We give approximation algorithms to estimate the contextuality number of a dataset, and evaluate these on test empirical models whose contextuality number we can verify analytically or numerically. Finally, we apply our algorithms to a cancer biomarker dataset, as a first demonstration of using strong $k$-contextuality to identify a problem that classical machine learning may struggle with. 

Building on existing connections between contextuality and quantum advantage \cite{gao2021enhancing, anschuetzgao2022, anschuetz_arbitrary_2024}, we introduced strong $k$‑contextuality as an indicator of generative modeling tasks which may exhibit a classical-quantum memory separation. Specifically, we proved that any classical hidden Markov model requires at least $k+1$ hidden variables to represent a $k$‑strongly contextual distribution within a finite relative entropy. When $k$ scales quickly with the problem size, learning tasks can become intractable for classical generative models, while quantum generative models may remain efficient. Crucially, our empirical results show a performance gap between quantum and classical models that grows with the estimated $k$, suggesting that strong $k$‑contextuality can be a useful indicator of when we might expect quantum advantage in generative learning, in both asymptotic and practical settings.

It is important to note that although strong $k$‑contextuality provably determines when distributions are costly for classical models to learn, our results do not prescribe a formula to build quantum models that could learn these distributions easily. While it is true that there exist provably resource-intractable generative tasks that a quantum model can learn \cite{gao2021enhancing}, there is no \textit{guarantee} that such a model always exists.  Indeed, there also exist many counterexamples of contextual systems with no efficient representations~\cite{cirel1980quantum,acin2015combinatorial}, quantum or classical. In other words, the fact that it is impossible to extend our techniques to an analogous quantum lower bound leaves room for more efficient quantum models to exist, but it is unclear if and how such quantum models can be constructed in general. Furthermore, whether the resource lower bound established using strong $k$-contextuality becomes prohibitive depends on the resource budget for that problem. At the same time, there are many other factors beyond contextuality that contribute to the cost of solving a problem. While strong $k$-contextuality provides a lower bound to the memory resources required, the actual resource cost could go well beyond that lower bound. 

We hope in future work to further our understanding of when contextuality can be used as a tool to guide the development of efficient quantum algorithms for classically hard problems. While separations that are provable and empirically verifiable are often found only in quantum-inspired distributions \cite{bouland2019complexity, anschuetz_arbitrary_2024}, our numerical results take a first step towards evidence that contextuality-related separations may exist in practically relevant datasets. It would be valuable to empirically identify additional contextual datasets to test the necessity of such intrinsically quantum connections, and study if and when strong-$k$ contextuality could predict performance in useful generative tasks. We also hope to explore the connections between strong $k$-contextuality and other measures of classical hardness. For instance, prior work has connected other measures of contextuality to the presence of magic or Wigner negativity in a state \cite{andersbrowne2009, raussendorfmbqc2013, abramsky2017contextualfraction,howard_contextuality_2014,bermejo-vega_contextuality_2017, emeriau_interplay_2022, booth2022continuous, raussendorf_role_2023}. It would be interesting to extend this analysis to strong $k$-contextuality and explore potential connections between strong $k$-contextuality and quantum advantage.

%Another promising direction is to treat $k$‑contextuality as a guide for algorithm development: through better understanding of why classical HMMs are deficient in representing highly contextual distributions and how quantum physics, [...], can be used to overcome this limitation, we may design contextuality-aware algorithms---quantum, classical, or hybrid---that targets contextuality-related challenges. Exploring such targeted algorithms, and seeing how much contextuality can explain or help to overcome the limit for classical generative models before hitting new bottlenecks, remains an exciting opportunity for both the field of learning theory and quantum advantage. \willers{this may be too much claim lol}

%While our metrics may reveal when a learning task may be less challenging for quantum generative models compared to their classical counterparts, they do not prescribe a quantum model in which learning the same data set is always easy. One may expect that as strong k contextuality points out what makes a distribution hard to learn, there may also be hopes of developing ml models -- either quantum or classical -- that address these specifically. 

\appendix
\subsection{\texorpdfstring{$k$}{k}-Contextuality Estimation as a Hypergraph Coloring Problem} \label{apx:coloring}
Let $S_i := S_i^{\left\{C_i\right\}}$, where the latter is as defined in Eq.~\eqref{eq:s_e_m_def}. Notice that $S_i \cap S_j\neq\varnothing$ implies that there is a common valid global value assignment between contexts $C_i$ and $C_j$, i.e., that $C_i$ and $C_j$ are compatible. 

Then the main insight that allows us to turn finding k-contextuality into a hypergraph coloring problem is the following lemma.

\begin{lemma} \label{lemma:sparsity}
    If $|S_i| \leq d$ for all $C_i \in \mathcal{M}$, then
    \begin{equation}
            \bigcap^{d+1}_{l}S_{i_l} \neq \emptyset \quad \forall 1 \leq i_1 < \ldots < i_{d+1} \leq m \iff\bigcap^{m}_{i}S_{i} \neq \emptyset.
    \end{equation}
\end{lemma}

i.e., if an empirical model is $d$-sparse (has fewer than $d$ consistent global assignments for every context), a group of $m$ contexts is compatible if and only if every $(d+1)$-subset of those contexts are compatible.

\paragraph{Proof (Lemma \ref{lemma:sparsity})}
Let us start by proving the $m=d+2$ case in the forward direction. Consider the bipartite graph with nodes $\{S_1,...,S_{d+2}\}$ and $\{s \in \mathcal{E}(X)\}$, and edges $E = \{(S_i, s) | s \in S_i\}$. 

Since $|S_i| \leq d$ for all $C_i \in \mathcal{M}$, we know deg($S_i$) $\leq d$. There are $(d+2)$ $S_i$ nodes, so there must be a maximum of $d(d+2)$ edges in the graph. We can use this to upper bound the number of global value assignments $s$ with deg($s$)$\geq d+1$:

\begin{equation} \label{eq:degfact}
    \#[\text{deg}(s) \geq d+1] \quad \leq\quad \frac{d(d+2)}{(d+1)}\quad <\quad d+2.
\end{equation}
Since $\bigcap^{d+1}_{l}S_{i_l} \neq \emptyset \quad \forall \quad 1 \leq i_1 < ... < i_{d+1} \leq d+2$, we know that for each of the $\binom{d+2}{d+1} = d+2$ subsets $\{i_l\}$ there must be some global assignment $s_{\{i_l\}}$ with deg($s_{\{i_l\}}$)$\geq d+1$.

Now, assume for the sake of contradiction that $\bigcap^{d+2}_{i}S_{i} = \emptyset$. This means that there is no global assignment $s$ with deg($s$)$\geq d+2$, meaning each of the $s_{\{i_l\}}$ must be unique. Since there are $d+2$ subsets $\{i_l\}$,$\#[\text{deg}(s) \geq d+1] \quad \geq \quad d+2.$ However, this is not possible by Eq. \ref{eq:degfact}, so our assumption must be wrong, and $\bigcap^{d+2}_{i}S_{i} \neq \emptyset$.

The backward direction is trivial. 

Thus, we have proved that if $|S_i| \leq d$ for all $C_i \in \mathcal{M}$, then \eqref{lemma:sparsity} is sastified for $m = d +2$. The rest follows inductively on $m$.

%\begin{equation}
%    \begin{aligned}
%        \bigcap^{d+1}_{l}&S_{i_l} \neq \emptyset \quad \forall \quad 1 \leq i_1 < %... < i_{d+1}\\
%        &\leq d+2 \qquad \Leftrightarrow \qquad \bigcap^{d+2}_{i}S_{i} \neq %\emptyset.
%    \end{aligned}
%\end{equation}

Using Lemma \ref{lemma:sparsity}, if we are able to draw a $(d+1)$-uniform compatibility hypergraph $\mathcal{G}$ (meaning edges represent the compatibility of $(d+1)$ contexts), the minimum number of cliques needed to cover $\mathcal{G}$ is exactly what we want---the minimum number of global value assignments we need to describe this empirical model. 

Since the dual problem of a clique cover problem is a coloring problem, we can equivalently phrase this as a hypergraph coloring problem.
\newpage
\section*{Acknowledgements}
This work is supported in part by Wellcome Leap as part of the `Quantum Biomaker Algorithms for Multimodal Cancer Data' research project within the the Quantum for Bio (Q4Bio) Program, in part by IBM Quantum under the IBM-UChicago Quantum Centric Supercomputing Collaboration (under agreement number MAS000364, with access to the fleet of IBM Quantum systems), in part by STAQ under award NSF Phy-1818914/232580; in part by the US Department of Energy Office of Advanced Scientific Computing Research, Accelerated Research for Quantum Computing Program; and in part by the NSF Quantum Leap Challenge Institute for Hybrid Quantum Architectures and Networks (NSF Award 2016136), in part based upon work supported by the U.S. Department of Energy, Office of Science, National Quantum Information Science Research Centers, and in part by the Army Research Office under Grant Number W911NF-23-1-0077. This work was completed in part with resources provided by the University of Chicago’s Research Computing Center. E.R.A. is funded in part by the Walter Burke Institute for Theoretical Physics at Caltech.

FTC is the Chief Scientist for Quantum Software at Infleqtion and an advisor to Quantum Circuits, Inc.

\bibliographystyle{IEEEtran}
\bibliography{main,other_papers}

% Generated by IEEEtran.bst, version: 1.14 (2015/08/26)
\begin{thebibliography}{10}
\providecommand{\url}[1]{#1}
\csname url@samestyle\endcsname
\providecommand{\newblock}{\relax}
\providecommand{\bibinfo}[2]{#2}
\providecommand{\BIBentrySTDinterwordspacing}{\spaceskip=0pt\relax}
\providecommand{\BIBentryALTinterwordstretchfactor}{4}
\providecommand{\BIBentryALTinterwordspacing}{\spaceskip=\fontdimen2\font plus
\BIBentryALTinterwordstretchfactor\fontdimen3\font minus \fontdimen4\font\relax}
\providecommand{\BIBforeignlanguage}[2]{{%
\expandafter\ifx\csname l@#1\endcsname\relax
\typeout{** WARNING: IEEEtran.bst: No hyphenation pattern has been}%
\typeout{** loaded for the language `#1'. Using the pattern for}%
\typeout{** the default language instead.}%
\else
\language=\csname l@#1\endcsname
\fi
#2}}
\providecommand{\BIBdecl}{\relax}
\BIBdecl

\bibitem{arute2019quantum}
F.~Arute \emph{et~al.}, ``Quantum supremacy using a programmable superconducting processor,'' \emph{Nature}, vol. 574, no. 7779, pp. 505--510, 2019.

\bibitem{ZHU2022240}
Q.~Zhu, S.~Cao, F.~Chen, M.-C. Chen, X.~Chen, T.-H. Chung, H.~Deng, Y.~Du, D.~Fan, M.~Gong \emph{et~al.}, ``Quantum computational advantage via 60-qubit 24-cycle random circuit sampling,'' \emph{Sci. Bull.}, vol.~67, no.~3, pp. 240--245, 2022.

\bibitem{doi:10.1126/science.abe8770}
H.-S. Zhong, H.~Wang, Y.-H. Deng, M.-C. Chen, L.-C. Peng, Y.-H. Luo, J.~Qin, D.~Wu, X.~Ding, Y.~Hu \emph{et~al.}, ``Quantum computational advantage using photons,'' \emph{Science}, vol. 370, no. 6523, pp. 1460--1463, 2020.

\bibitem{gao2021enhancing}
X.~Gao, E.~R. Anschuetz, S.-T. Wang, J.~I. Cirac, and M.~D. Lukin, ``Enhancing generative models via quantum correlations,'' \emph{Phys. Rev. X}, vol.~12, p. 021037, 2022.

\bibitem{anschuetzgao2022}
E.~R. Anschuetz, H.-Y. Hu, J.-L. Huang, and X.~Gao, ``Interpretable quantum advantage in neural sequence learning,'' \emph{PRX Quantum}, vol.~4, p. 020338, 2023.

\bibitem{anschuetz_arbitrary_2024}
\BIBentryALTinterwordspacing
E.~R. Anschuetz and X.~Gao, ``Arbitrary {Polynomial} {Separations} in {Trainable} {Quantum} {Machine} {Learning},'' Feb. 2024, arXiv:2402.08606 [quant-ph]. [Online]. Available: \url{http://arxiv.org/abs/2402.08606}
\BIBentrySTDinterwordspacing

\bibitem{Abramsky_2011}
S.~Abramsky and A.~Brandenburger, ``The sheaf-theoretic structure of non-locality and contextuality,'' \emph{New J. Phys.}, vol.~13, no.~11, p. 113036, nov 2011.

\bibitem{pearl1985bayesian}
\BIBentryALTinterwordspacing
J.~Pearl, ``Bayesian networks: A model of self-activated memory for evidential reasoning,'' in \emph{Proceedings of the 7th Conference of the Cognitive Science Society, University of California, Irvine, CA, USA}, 1985, pp. 329--334. [Online]. Available: \url{https://cognitivesciencesociety.org/wp-content/uploads/2019/01/cogsci_7.pdf}
\BIBentrySTDinterwordspacing

\bibitem{10.1214/aoms/1177699147}
L.~E. Baum and T.~Petrie, ``{Statistical Inference for Probabilistic Functions of Finite State Markov Chains},'' \emph{Ann. Math. Stat.}, vol.~37, no.~6, pp. 1554--1563, 1966.

\bibitem{Hopfield2554}
J.~J. Hopfield, ``Neural networks and physical systems with emergent collective computational abilities,'' \emph{Proc. Natl. Acad. Sci. U.S.A.}, vol.~79, no.~8, pp. 2554--2558, 1982.

\bibitem{10.1162/neco.1997.9.8.1735}
S.~Hochreiter and J.~Schmidhuber, ``Long short-term memory,'' \emph{Neural Comput.}, vol.~9, no.~8, pp. 1735--1780, 1997.

\bibitem{10.5555/3295222.3295349}
\BIBentryALTinterwordspacing
A.~Vaswani, N.~Shazeer, N.~Parmar, J.~Uszkoreit, L.~Jones, A.~N. Gomez, L.~Kaiser, and I.~Polosukhin, ``Attention is all you need,'' in \emph{Proceedings of the 31st International Conference on Neural Information Processing Systems}, ser. NIPS'17, I.~Guyon, U.~V. Luxburg, S.~Bengio, H.~Wallach, R.~Fergus, S.~Vishwanathan, and R.~Garnett, Eds.\hskip 1em plus 0.5em minus 0.4em\relax Red Hook, NY, USA: Curran Associates, Inc., 2017, pp. 6000--6010. [Online]. Available: \url{https://papers.nips.cc/paper_files/paper/2017/hash/3f5ee243547dee91fbd053c1c4a845aa-Abstract.html}
\BIBentrySTDinterwordspacing

\bibitem{9635256}
M.~Baucum, A.~Khojandi, and T.~Papamarkou, ``Hidden {Markov} models as recurrent neural networks: An application to {Alzheimer's} disease,'' in \emph{2021 IEEE 21st International Conference on Bioinformatics and Bioengineering (BIBE)}.\hskip 1em plus 0.5em minus 0.4em\relax Los Alamitos, CA, USA: IEEE Computer Society, Oct. 2021, pp. 1--6.

\bibitem{acin2015combinatorial}
A.~Ac{\'\i}n, T.~Fritz, A.~Leverrier, and A.~B. Sainz, ``A combinatorial approach to nonlocality and contextuality,'' \emph{Commun. Math. Phys.}, vol. 334, pp. 533--628, 2015.

\bibitem{PhysRevLett.115.150401}
J.~V. Kujala, E.~N. Dzhafarov, and J.-A. Larsson, ``Necessary and sufficient conditions for an extended noncontextuality in a broad class of quantum mechanical systems,'' \emph{Phys. Rev. Lett.}, vol. 115, p. 150401, Oct 2015.

\bibitem{plociennik2008approximation}
K.~Plociennik, ``Approximating independent set and coloring in random uniform hypergraphs,'' in \emph{Mathematical Foundations of Computer Science 2008}, E.~Ochma{\'{n}}ski and J.~Tyszkiewicz, Eds.\hskip 1em plus 0.5em minus 0.4em\relax Berlin, Heidelberg: Springer Berlin Heidelberg, 2008, pp. 539--550.

\bibitem{PhysRevA.89.062315}
G.~H. Low, T.~J. Yoder, and I.~L. Chuang, ``Quantum inference on {Bayesian} networks,'' \emph{Phys. Rev. A}, vol.~89, p. 062315, Jun. 2014.

\bibitem{baum1970maximization}
L.~E. Baum, T.~Petrie, G.~Soules, and N.~Weiss, ``A maximization technique occurring in the statistical analysis of probabilistic functions of markov chains,'' \emph{Ann. Math. Statist.}, vol.~41, no.~1, pp. 164--171, 1970.

\bibitem{itensor}
\BIBentryALTinterwordspacing
M.~Fishman, S.~R. White, and E.~M. Stoudenmire, ``The {ITensor} software library for tensor network calculations,'' \emph{SciPost Phys. Codebases}, p.~4, 2022. [Online]. Available: \url{https://scipost.org/10.21468/SciPostPhysCodeb.4}
\BIBentrySTDinterwordspacing

\bibitem{harley1990uci}
C.~Harley, R.~Reynolds, and M.~Noordewier, ``{Molecular Biology (Promoter Gene Sequences)},'' UCI Machine Learning Repository, 1990, {DOI}: https://doi.org/10.24432/C5S01D.

\bibitem{sanyal2012long}
A.~Sanyal, B.~R. Lajoie, G.~Jain, and J.~Dekker, ``The long-range interaction landscape of gene promoters,'' \emph{Nature}, vol. 489, no. 7414, pp. 109--113, 2012.

\bibitem{le2019classifying}
N.~Q.~K. Le, E.~K.~Y. Yapp, N.~Nagasundaram, and H.-Y. Yeh, ``Classifying promoters by interpreting the hidden information of dna sequences via deep learning and combination of continuous fasttext n-grams,'' \emph{Frontiers in bioengineering and biotechnology}, vol.~7, p. 305, 2019.

\bibitem{cirel1980quantum}
B.~S. Cirel'son, ``Quantum generalizations of {Bell}'s inequality,'' \emph{Lett. Math. Phys.}, vol.~4, pp. 93--100, 1980.

\bibitem{bouland2019complexity}
A.~Bouland, B.~Fefferman, C.~Nirkhe, and U.~Vazirani, ``On the complexity and verification of quantum random circuit sampling,'' \emph{Nat. Phys.}, vol.~15, no.~2, pp. 159--163, 2019.

\bibitem{andersbrowne2009}
\BIBentryALTinterwordspacing
J.~Anders and D.~E. Browne, ``Computational power of correlations,'' \emph{Phys. Rev. Lett.}, vol. 102, p. 050502, Feb 2009. [Online]. Available: \url{https://link.aps.org/doi/10.1103/PhysRevLett.102.050502}
\BIBentrySTDinterwordspacing

\bibitem{raussendorfmbqc2013}
\BIBentryALTinterwordspacing
R.~Raussendorf, ``Contextuality in measurement-based quantum computation,'' \emph{Phys. Rev. A}, vol.~88, p. 022322, Aug 2013. [Online]. Available: \url{https://link.aps.org/doi/10.1103/PhysRevA.88.022322}
\BIBentrySTDinterwordspacing

\bibitem{abramsky2017contextualfraction}
\BIBentryALTinterwordspacing
S.~Abramsky, R.~S. Barbosa, and S.~Mansfield, ``Contextual fraction as a measure of contextuality,'' \emph{Phys. Rev. Lett.}, vol. 119, p. 050504, Aug 2017. [Online]. Available: \url{https://link.aps.org/doi/10.1103/PhysRevLett.119.050504}
\BIBentrySTDinterwordspacing

\bibitem{howard_contextuality_2014}
\BIBentryALTinterwordspacing
M.~Howard, J.~Wallman, V.~Veitch, and J.~Emerson, ``\BIBforeignlanguage{en}{Contextuality supplies the ‘magic’ for quantum computation},'' \emph{\BIBforeignlanguage{en}{Nature}}, vol. 510, no. 7505, pp. 351--355, Jun. 2014. [Online]. Available: \url{https://www.nature.com/articles/nature13460}
\BIBentrySTDinterwordspacing

\bibitem{bermejo-vega_contextuality_2017}
\BIBentryALTinterwordspacing
J.~Bermejo-Vega, N.~Delfosse, D.~E. Browne, C.~Okay, and R.~Raussendorf, ``\BIBforeignlanguage{en}{Contextuality as a {Resource} for {Models} of {Quantum} {Computation} with {Qubits}},'' \emph{\BIBforeignlanguage{en}{Physical Review Letters}}, vol. 119, no.~12, p. 120505, Sep. 2017. [Online]. Available: \url{https://link.aps.org/doi/10.1103/PhysRevLett.119.120505}
\BIBentrySTDinterwordspacing

\bibitem{emeriau_interplay_2022}
\BIBentryALTinterwordspacing
P.-E. Emeriau, ``The {Interplay} between {Quantum} {Contextuality} and {Wigner} {Negativity},'' Apr. 2022, arXiv:2204.08782 [quant-ph]. [Online]. Available: \url{http://arxiv.org/abs/2204.08782}
\BIBentrySTDinterwordspacing

\bibitem{booth2022continuous}
R.~I. Booth, U.~Chabaud, and P.-E. Emeriau, ``Contextuality and {Wigner} negativity are equivalent for continuous-variable quantum measurements,'' \emph{Phys. Rev. Lett.}, vol. 129, p. 230401, 2022.

\bibitem{raussendorf_role_2023}
\BIBentryALTinterwordspacing
R.~Raussendorf, C.~Okay, M.~Zurel, and P.~Feldmann, ``The role of cohomology in quantum computation with magic states,'' \emph{Quantum}, vol.~7, p. 979, Apr. 2023, arXiv:2110.11631 [quant-ph]. [Online]. Available: \url{http://arxiv.org/abs/2110.11631}
\BIBentrySTDinterwordspacing

\end{thebibliography}

\end{document}